\documentclass[fdp,fleqn]{w-art}
\usepackage{times}
\usepackage{w-thm}
\usepackage{eucal}
\usepackage[latin1]{inputenc}
\usepackage[english]{babel}
\usepackage{amsmath}
\usepackage{wasysym}
\usepackage{amsfonts}
\usepackage{amssymb}
\usepackage[dvips]{graphicx}
\usepackage{pifont}
\usepackage[]{graphicx}

\newcommand{\ba}{\begin{align}}
\newcommand{\ena}{\end{align}}
\newcommand{\be}{\begin{equation}}
\newcommand{\ee}{\end{equation}}
\newcommand{\fr}{\frac}
\newcommand{\ft}{\footnote}

\def\a{\alpha}
\def\b{\beta}
\def\g{\gamma}

\def\d{\delta}

\def\h{\eta}

\def\l{\lambda}
\def\L{\Lambda}
\def\m{\mu}
\def\n{\nu}

\def\r{\rho}

\def\vf{\varphi}

\def\cB{{\cal B}}
\def\cE{{\cal E}}

\def\cJ{{\cal J}}

\def\cL{{\cal L}}

\def\cF{{\cal F}}

\def\cA{{\cal A}}

\def\cR{{\cal R}}

\def\cA{{\cal A}}

\def\pe{\prime}
\def\eq{\equiv}

\def\pr{\partial}
\def\prd{\partial \cdot}

\def\12{\frac{1}{2}}

\begin{document}
\DOIsuffix{theDOIsuffix}
\Volume{55}
\Issue{1}
\Month{01}
\Year{2007}



\title[Geometric massive higher spins]{Geometric massive higher spins and current exchanges}


\author[D. Francia]{Dario Francia 
  \footnote{ E-mail:~\textsf{francia@chalmers.se}.
}}
\address[]{Chalmers University of Technology, \\
Department of Fundamental Physics,
S-41296 G\"oteborg - Sweden}
\begin{abstract}
Generalised Fierz-Pauli mass terms allow to describe
massive higher-spin fields on flat background by means of 
simple quadratic deformations
of the corresponding geometric, massless Lagrangians.
In this framework there is no need for auxiliary fields.
We briefly review the construction in the bosonic case and 
study the interaction of these massive fields with external sources,
computing the corresponding propagators. 
In the same fashion as for the massive graviton, but differently from theories
where auxiliary fields are present, the structure of the current exchange  is completely determined
by the form of the mass term itself.
\vskip 20pt
\noindent
\emph{Based on the talk presented at the $3^{rd}$ RTN-Forces Universe Workshop - Valencia, Spain, 1-5 October 2007} 
\end{abstract}

\maketitle                   





\section{Introduction and summary}

In this contribution we would like to briefly review the work
\cite{dario07}, where we proposed massive Lagrangians for higher-spin 
fields  from a perspective such that, 
in particular, no need for auxiliary fields emerges. In addition, we compute here the
propagator of those theories, along the lines of the extensive
analysis of  similar issues performed in \cite{FMS} and \cite{fms08}.

 As an introduction, let us   
recall that  the Lagrangian description of massive, lower-spin 
fields\ft{i.e. fields with spin $s \leq 2$.}
is both simple and unique. For instance, in the spin $2$ case,
the Fierz-Pauli Lagrangian
\cite{fierzpauli},
\be \label{FPspin2}
\cL \, = \, \12 \, h^{\, \m \n}\, \{\cR_{\m \n} \, - \, \12 \, \h_{\m \n} \, \cR \, - \, m^{\, 2} \, 
(h_{\m \n} \, - \, \h_{\m \n}\, h^{\, \a}_{\, \ \a})\} \, ,
\ee
where $\cR_{\m \n}$ and $\cR$ indicate the linearised
Ricci tensor and Ricci scalar respectively, 
describes  the  \emph{only}  consistent quadratic deformation
of the linearised Einstein-Hilbert theory \cite{Boulware:1973my}. 
Actually, for all bosonic and fermionic lower-spin fields, 
a consistent and unique massive theory can be obtained adding 
to their massless Lagrangians  suitable quadratic terms.

 By contrast, the traditional description
of massive higher-spin fields\ft{For reviews on the subject 
of higher-spin gauge fields  see \cite{review}.}
is neither as simple, nor it is unique, and in particular, 
for spin $s \geq \fr{5}{2}$,
auxiliary fields are usually found to be needed off-shell. 
As already noticed in \cite{fierzpauli} they 
can be chosen in several ways, so that the theory 
looses uniqueness (although
a minimal choice was first identified in 
\cite{sh}\ft{To describe spin-$s$ massive 
degrees of freedom in \cite{sh} a symmetric, traceless rank-$s$ tensor was introduced,
together with a set of symmetric and traceless auxiliary
tensors of rank $s-2, s-3, \dots, 0$. In fact, as already
noticed in \cite{singh}, by means of  suitable field
redefinitions it is possible to collect all those fields in a set of only 
\emph{two, traceful} symmetric tensors 
of rank $s$ and $s-3$ respectively. }  ), while
the resulting Lagrangians do not look like simple 
quadratic deformations of the corresponding massless ones.

 The key observation in order to understand the origin of this difference
between lower- and higher-spin fields is to notice that, still focusing on the example of spin $2$, 
consistency of the dynamics described by (\ref{FPspin2}) is guaranteed by the Bianchi identity
satisfied by the Einstein tensor,
\be
\pr^{\, \a} \{\cR_{\a \m}  -  \12 \, \h_{\a \m} \, \cR\} \,\eq \, 0 \, ,
\ee
which, in its turn, reflects the geometrical underpinnings of the
massless sector of the theory. Implementing this piece of information on-shell
allows to recover the Fierz-Pauli constraint
\be \label{fpconstr}
\pr^{\, \a} \,  h_{\, \a \, \m} \, - \, \pr_{\, \m} \, h^{\, \a}_{\, \ \a} \, = \, 0 \, ,
\ee
which is ultimately responsible for the reduction of the equations
of motion to the system 
\be \label{fierzsyst}
\begin{split}
&(\Box \, - \, m^2) \, h_{\, \m \n} \, = \, 0 \ , \\
&\pr^{\, \a} \, h_{\, \a \m} \, = \, 0 \, , \\
&h^{\, \a}_{\, \ \a} \, = \, 0 \, ,
\end{split}
\ee
describing the
irreducible propagation of massive, spin-$2$ degrees of freedom.
On the other hand, in the (Fang-)Fronsdal theory of massless higher-spin fields
\cite{fronsdal} the corresponding Einstein tensors 
\emph{are not fully divergenceless}, as a reflection
of their lack of direct geometrical meaning. 
In fact, whereas the higher-spin curvatures
introduced by de Wit and Freedman in \cite{dwf}
are \emph{fully} gauge-invariant under the abelian
gauge transformation of the potential\ft{Dots indicate symmetrization
of the $s$ indices.},
\be \label{gauge}
\d \, \vf_{\m_1 \, \dots \, \m_s} \, = \, \pr_{\, \m_1} \, \L_{\m_2 \, \dots \, \m_s}\, +
\, \dots \, ,
\ee
without any conditions on $\L_{\m_1 \, \dots \, \m_{s-1}}$ , 
it turns out that the basic equation of the Fronsdal theory,
\be  \label{fronsdalT}
\cF_{\m_1 \, \dots \, \m_s} \, \equiv \, \Box \, \vf_{\m_1 \, \dots \m_s}\, -  \, 
\pr_{\m_1} \, \pr^{\, \a} \,  \vf_{\a \, \m_2 \, \dots \, \m_s} \, + \, \dots \, + \, 
\pr_{\m_1}\pr_{\m_2}\, \vf_{\, \ \a \,\m_3 \, \dots \m_s}^{\, \a}
+ \, \dots \, = \, 0 \, ,
\ee
is gauge-invariant under (\ref{gauge}) only if the 
parameter is taken to be \emph{traceless}: 
$\L^{\a}_{\, \ \a \,\m_3 \, \dots \m_{s - 1}} \equiv 0$, 
a condition which indicates that such a theory as it stands
cannot have a direct geometrical 
interpretation.
Moreover, as a consequence of the constraint on the trace
of the gauge parameter, the 
corresponding Einstein tensor need not be
(and is not) identically divergenceless, and for
this reason, as discussed in \cite{ady87}, in order to 
get a consistent massive theory, a simple quadratic deformation of
the Fronsdal Lagrangians alone is not enough, and auxiliary fields 
are to be introduced\ft{See \cite{buch} 
for a more recent approach.}.

 A \emph{geometric} description of massless higher-spin gauge fields, 
where all quantities of dynamical interest are actually built
from curvatures, was proposed in \cite{fs, fs2, FMS} for the case
of symmetric tensors\ft{Generalisations to the case
of mixed-symmetry gauge fields have been given in
\cite{BekBoulDem}.}. The main outcome of the full construction 
is that, out of infinitely many geometric Lagrangians
available at the free level, consistency with the
coupling to an external source requires the 
theory to have a \emph{unique} form.
In particular the Einstein tensor of this theory,
in the compact notation\ft{All symmetrised indices are implicit, 
and  symmetrization without factors among indices is 
always understood in the product of different tensors. $\h$ is the 
``mostly-plus'' space-time metric in $d$ dimensions,  
``primes'', as well as numbers in 
square brackets, denote traces while divergences are denoted by
``$\prd$''. Useful combinatorial identities are \begin{alignat}{4}
\left( \pr^{\, p} \vf  \right)^{\, \pe} \, & =  \, \Box 
 \pr^{\, p-2}  \vf  +  2  \pr^{\, p-1}   \prd \vf +  \pr^{\, p} 
\vf^{\, \pe} \, , & \qquad  \ \ \ \  \partial^{\, p}  \partial^{\, q}  \, & = \, {p+q \choose p} 
\partial^{\, p+q} \, , \notag  \\
\left( \eta^k  \vf    \right)^{\, \pe} \, & = \, \left[D
 +  2 (s+k-1)   \right] \eta^{\, k-1}  \vf + \eta^k
\vf^{\, \pe} \, , & \qquad \ \ \ \ \eta  \eta^{\, n-1} \, & =   \, n  \eta^{\, n} \, . \notag 
\end{alignat}.}
of those works, that we shall also
exploit here, can be written 
\be \label{einst}
\cE_{\, \vf} \, = \, \cA_{\, \vf} \, - \, \12 \, \h \, \cA_{\, \vf}^{\, \pe}
\, + \, \h^2 \, \cB_{\, \vf} \, .
\ee
The generalised Ricci tensor $\cA_{\, \vf}$, which is fully
constructed out of curvatures \cite{FMS}, 
admits a particularly simple interpretation when written in terms
of the Fronsdal tensor $\cF$:
\be \label{a}
\cA_{\, \vf} \, = \, \cF \, - \, 3 \, \pr^3 \g_{\, \vf} \, ,
\ee
where $\g_{\, \vf}$ is a \emph{non-local} 
tensor\ft{Since for spin $s$ the curvatures of \cite{dwf}
contain $s$ derivatives, in order for the differential operator
appearing in (\ref{a})
to carry the same dimensions 
as the D'Alembertian operator, non-localities
are to be introduced as an unavoidable intermediate feature
of the geometric construction.}, transforming under (\ref{gauge}) with the trace of the 
gauge parameter:
\be
\d \, \g_{\, \vf} \, = \, \L^{\, \pe} \, .
\ee
This implies that, with the same gauge fixing, it is possible
both to remove all non localities from the equations of motion
and to recover the Fronsdal form (\ref{fronsdalT})\ft{This comes from the fact
that the Lagrangian equations $\cE_{\, \vf} = 0$
can be shown to imply $\cA_{\, \vf} = 0$.}, thus showing the consistency 
of the construction.
It should be stressed that all 
non-localities can be removed also \emph{off-shell}, and without performing
any gauge-fixing, at the price of introducing \emph{auxiliary fields}.
Indeed, whereas an unconstrained description of the Fronsdal dynamics
involving auxiliary fields was already known since some time \cite{pt}, the form 
(\ref{a}) of the tensor $\cA_{\, \vf}$ suggests
a very economical alternative option, first proposed in \cite{fs2, st}, 
that is to substitute the non-local tensor $\g_{\, \vf}$
with a \emph{compensator} field $\a$ having the same gauge
transformation:
\be
\d \, \a \, = \, \L^{\, \pe} \, .
\ee
In this way the \emph{local} tensor 
\be \label{aloc}
\cA \, = \, \cF \, - \, 3 \, \pr^3 \, \a \, ,
\ee
can be used as a starting point to build  local, 
unconstrained Lagrangians \cite{fs3}, whose completion
only requires a further auxiliary field $\b$, with 
the transformation property $\d \b = \prd \prd \prd \L$.
Finally, the elimination of the higher derivatives appearing in 
connection with the compensator $\a$ can again be implemented
in a rather economical fashion, thus leading to
an ordinary derivative unconstrained Lagrangian involving
a total of five fields for any spin, as described in 
\cite{dario07} (see also \cite{BuchQuartet} for related work).

 For our present purposes, the main feature of the tensor (\ref{einst}) (and actually of 
all tensors among the infinitely many available at the free level) is to be
\emph{identically divergenceless},
as required by the absence of constraints on the gauge parameter.
It is then possible to
look for quadratic deformations of the corresponding class of Lagrangians
in the spirit of the Fierz-Pauli description of the massive graviton
leading to  (\ref{FPspin2}). The concrete realization of this 
program was the main result of
\cite{dario07} where 
generalised Fierz-Pauli mass terms were proposed, both for bosons and fermions of any spin.
Their form, for the simple cases  of spin $4$ and spin $\fr{7}{2}$,
looks
\be
M_{\, \vf} =  \vf_{\, \m_1 \dots \m_4} \, - \, (\h_{\, \m_1 \m_2} \, 
\vf^{\, \a}_{\ \ \a \, \m_3 \m_4} \, + \, \dots\,) \, 
- \, (\h_{\, \m_1 \m_2} \, \h_{\, \m_3 \m_4} \, + \, \dots \, )\, 
\vf^{\, \a \,  \b}_{\ \ \a \, \b} \, , 
\ee
\be
 M_{\, \psi}  =  \psi_{\, \m_1 \m_2 \m_3}  - 
(\g_{\, \m_1} \g^{\, \a}  \psi_{\, \a \m_2 \m_4} +  \dots) 
-    (\h_{\, \m_1 \m_2}  
\psi^{\, \a}_{\ \ \a \, \m_3} \, + \, \dots\,) 
-  (\g_{\, \m_1} \, \h_{\, \m_2 \m_3}  \, + \, \dots \, )
\g^{\, \a}  \psi^{\, \b}_{\ \ \a \, \b}  , 
\ee
where it is already possible to appreciate one basic feature of the general
result: all (gamma-)traces of the field enter the mass terms, 
in a sequence starting with the Fierz-Pauli contributions.
We shall review the derivation of the generalised Fierz-Pauli mass terms
in Section \ref{sec2}, restricting our attention to bosons.

 In \cite{FMS, fms08} a systematic analysis of the interaction
between unconstrained higher-spin fields and external sources, 
both in flat and in (A)dS backgrounds, was proposed. 
This provided first of all a test of the consistency
of the unconstrained Lagrangians, both in the geometric, non-local
framework of \cite{fs, fs2, FMS} and in its minimal local counterpart of \cite{fs3,
FMS}, \emph{versus} the structure of the corresponding propagators.
Together with that, a few interesting issues were also addressed, such as the
persistence of the vDVZ discontinuity \cite{vdvz} for all spins on flat backgrounds
\cite{FMS, fms08}, 
and its disappearance on (A)dS spaces \cite{fms08}. 
 
 Here we would like to supplement the results obtained in 
\cite{FMS, fms08}, computing the current exchange for 
the massive Lagrangians of \cite{dario07}, whose consistency
can then be compared with the result obtained in the local
setting by Kaluza-Klein reduction in 
\cite{FMS, fms08}. This is done in Section \ref{sec3}.

 An interesting aspect of this analysis relies in the fact that, 
under the conditions of conservation of the sources, 
\emph{the full structure of the massive propagator is
encoded in the coefficients of the mass terms}.
This is already true for the massive graviton, 
where it represents a clue to understand the
``rigidity'' of the Fierz-Pauli mass term of (\ref{FPspin2}), while in our 
framework this feature provides a rather strong
consistency check on the form of the generalised Fierz-Pauli mass term
proposed in \cite{dario07}.

\section{Generalised Fierz-Pauli mass terms} \label{sec2}

 In this Section we would like to summarise the construction
of the generalised Fierz-Pauli mass terms given in \cite{dario07}, 
focusing for simplicity on the bosonic case\ft{For the corresponding
discussion for fermions, together with an account
of the fermionic geometry underlying their massless
Lagrangians see \cite{dario07}.}.
For definitiness, 
we assume that the massless sector of the theory is 
described by the divergenceless Einstein tensor (\ref{einst}), with the
generalised Ricci tensor $\cA_{\, \vf}$ given by 
(\ref{a})\ft{Let us stress that the non local compensator tensor
$\g_{\, \vf}$ in (\ref{a}) must have a very specific 
form. In fact, as shown in \cite{FMS}, there are infinitely many
non local tensors displaying the same gauge transformation
as $\g_{\, \vf}$, each associated to a theory with the correct
classical behaviour at free level, and in particular to 
a divergenceless Einstein tensor, but 
possessing in general the wrong propagator. 
What allows to select the (unique) correct form of $\g_{\, \vf}$
is either the request that  the identity
\be
\prd \cA_{\, \vf} \, - \, \12 \pr \, \cA_{\, \vf}^{\, \pe} \, \equiv \, 0 \, ,
\ee
(crucial to ensure that the massless propagator have
the correct structure) be satisfied, or 
-equivalently- that the tensor $\cA_{\, \vf}$ be
identically doubly traceless.}.
We thus look for a massive Lagrangian
for higher-spin bosons of the form
\be \label{geomlagmass}
\cL \, = \,  \12 \, \vf \, \{\cE_{\, \vf} - \, m^{\, 2} \, M_{\, \vf}\}  \, ,
\ee
where $M_{\, \vf}$ is a linear
function of $\vf$ to be determined.
The general idea is that $M_{\, \vf}$ should be a
combination of \emph{all} the traces of $\vf$, as expected
in our unconstrained setting. Moreover, 
given that the divergence of the equation of motion
\be
\prd \, \{\cE_{\, \vf} - \, m^{\, 2} \, M_{\, \vf}\} \, = \, 0 \, ,
\ee
reduces to
\be \label{divM}
\prd M_{\, \vf} \, = \, 0 \, ,
\ee
it is clear that the issue at stake
is to understand what conditions
should be deduced from (\ref{divM}).
From the conceptual viewpoint, the main result of 
\cite{dario07} was to prove that  (\ref{divM})
should imply for all spins  the
Fierz-Pauli constraint
\be \label{fpagain}
\prd \vf \, - \, \pr \, \vf^{\, \pe} \, = \, 0 \, ,
\ee
since this condition reveals itself to be necessary and sufficient
to recover the irreducibility conditions (\ref{fierzsyst}), 
generalised to a rank-$s$ tensor.
To have an idea of how the procedure works, let us discuss in some detail,
for the example of spin $4$, both the relevance of (\ref{fpagain}) 
and  the corresponding solution for $M_{\, \vf}$.
We shall then show how to compute  $M_{\, \vf}$
for the spin-$s$ case,
exploiting the requirement  that (\ref{divM}) imply (\ref{fpagain}).

 \subsection{Spin $4$}

Let us consider the Lagrangian,
\be
\cL \, = \, \12 \, \vf \, \{\cA_{\vf}  \, - \, \12 \, \h \, \cA_{\vf}^{\, \pe} \, 
+ \, \cB_{\, \vf} \,- \, m^2 \, M_{\, \vf}\} \, ,
\ee 
where the explicit form of  $\cA_{\, \vf}$ in terms of the Fronsdal tensor
$\cF$ defined in  (\ref{fronsdalT})
is given by
\be
\begin{split}
& \cA_{\, \vf} \, = \, \cF \, - \, 3 \, \pr^{\, 3} \, \g_{\, \vf} \, , \\
& \g_{\, \vf} \, = \, \fr{1}{3\, \Box^{\, 2}} \prd \cF^{\, \pe}  \, - \, 
\fr{1}{3}\, \fr{\pr}{\Box^{\, 3}} \prd \prd \cF^{\, \pe} \, + \, 
\fr{1}{12}\, \fr{\pr}{\Box^{\, 2}}\,\cF^{\, \pe \pe} \, , 
\end{split}
\ee
while $\cB_{\, \vf}$ is fixed by the requirement that  $\prd \cE_{\, \vf} = 0$,
implying
\be
 \pr \, \cB_{\, \vf} \,  = \, \12 \, \prd \cA_{\, \vf}^{\, \pe} \, .
\ee
For the mass term $M_{\, \vf} $ we choose the general combination
\be
M_{\, \vf} \, = \, \vf \, + \, a \, \h \, \vf^{\, \pe} \, + b \, \h^{\, 2} \, \vf^{\, \pe \pe}\, .
\ee
where the coefficients $a$ and $b$  should be chosen in  such a way to 
(\ref{divM}) imply
$\cA_{\vf}^{\, \pe}\, = \, 0 $.
On the other hand,  $\cA_{\vf}^{\, \pe}$ starts with $\cF^{\, \pe}$ together with
terms containing at least one divergence of $\cF^{\, \pe}$, 
and from the explicit form of $\cF^{\, \pe}$
\be \label{Fprime}
\cF^{\, \pe} \, = \, 2 \, \Box \, \vf^{\, \pe} \, - \, 2 \, \prd \prd \vf \, + \, \pr \,
\prd \vf^{\, \pe} \, + \, \pr^{\, 2} \, \vf^{\, \pe \pe} \, ,
\ee
we see that the first two terms cannot be compensated by anything in
the remainder of $\cA_{\vf}^{\, \pe}$, unless
the combination
\be
\Box \, \vf^{\, \pe} \, -  \, \prd \prd \vf \, 
\ee
result  to be expressible in terms
of higher traces and divergences of $\vf$,  
as a consequence of the equations of motion.
This kind of condition is indeed implemented by the Fierz-Pauli constraint, 
but would not hold if we had a more general condition of the form 
$\prd \vf - k \, \pr \vf^{\, \pe} = 0$, 
with $k \neq 1$, thus showing the very peculiar role played by 
(\ref{fpagain}).
If we then assume to have fixed the coefficients 
$a$ and $b$ so that $\prd M_{\, \vf} \, = \, 0$ implies (\ref{fpagain}), 
it is possible to show that the 
following consequences  hold:
\be
\begin{split} 
& \cF \, = \, \Box \, \vf \, - \, \pr^{\, 2} \vf^{\, \pe} \, , \\
&\cF^{\, \pe} \, = 3 \, \pr^{\, 2} \, \vf^{\, \pe \pe}\, ,
\end{split}
\ee
which, in their turn, can be shown to imply $\cA_{\vf}^{\, \pe}\, = 0$.
Consequently, the
Lagrangian equation reduces on-shell to the form
\be \label{spin4}
\cA_{\, \vf} \, - \, m^{\, 2} \, M_{\, \vf} \, = \, 0 \, ,
\ee
where the proper solution for $M_{\, \vf}$
such as to guarantee that $\prd M_{\, \vf} = 0$ imply (\ref{fpagain}), together
with its consistency condition $\prd \vf^{\, \pe} \, = \, - \, \pr \, \vf^{\, \pe \pe}$
is
\be
M_{\, \vf} \, = \, \vf \, - \, \h \, \vf^{\, \pe} \, -  \, \h^{\, 2} \, \vf^{\, \pe \pe}\, .
\ee
From the double trace of (\ref{spin4}) one obtains  $\vf^{\, \pe \pe} =  0$,  which
implies $\vf^{\, \pe} =  0$ and finally
$(\Box \, - \, m^2) \, \vf = 0$, as required.

 \subsection{Spin $s$} \label{sec2.2}

 In the general case, as already stressed,
all traces of $\vf$ are expected to contribute
to $M_{\vf}$, so that, for $s = 2\,n$ or $s = 2\, n + 1$, its
general form would be 
\be \label{masstermb}
M_{\vf} \, =\, \vf \, + \, b_1 \, \h \, \vf^{\, \pe} \, + b_2 \, \h^{\, 2} \, \vf^{\, \pe \pe} \, 
+ \dots \, + \, b_k \, \h^{\, k}  \, \vf^{\, [k]} \, +  \dots  + \, b_n \, \h^{\, n} \,  \vf^{\, [n]} \, ,
\ee
where $n \, = \, [\fr{s}{2}]$. The same argument seen for spin $4$ applies also in this case:
we would like to obtain $\cA_{\vf}^{\, \pe} = 0$  as a consequence of
$\prd M_{\, \vf} \, = \, 0$. 
To this end we look for coefficents $b_1, \, \dots \, b_n$ such
that the divergence of (\ref{masstermb}) imply (\ref{fpagain})
\emph{together with its consistency conditions}
\be \label{conscond}
\prd \vf^{\, [k]} \, = \, - \, \fr{1}{2\, k \, - \, 1} \, \pr \, \vf^{\, [k + 1]} \, ,
\hspace{2cm} 
k \, = \, 1 \, \dots \, n \, ,
\ee
since it is possible to show that if the latter equations are
satisfied then $\cA_{\vf}^{\, \pe} = 0$.
More explicitly, 
if we write the divergence of $M_{\vf}$ in the form
\be \label{divMb}
\prd M_{\vf} \, = \, \prd \vf \, + \, b_1 \, \pr \, \vf^{\, \pe} \, + \, 
\dots \, + \, \h^{\, k} \, (b_k \, \prd \vf^{\, [k]} \, + \, b_{k + 1} \, \pr \vf^{\, [k + 1]}) \, +  \, \dots \, ,
\ee
and we define $\m_{\, \vf} \, \eq \, \prd \vf \, - \, \pr \, \vf^{\, \pe}$, 
then we would like to rearrange (\ref{divMb}) as
\be \label{diverM}
\prd M_{\vf} \, = \, \m_{\vf} \, + \, \l_1 \, \h \, \m_{\vf}^{\, \pe} \, +
\, \dots \, + \, \l_k \, \h^{\, [k]}\, \m_{\vf}^{\, [k]} \, + \, \dots \, .
\ee
In this fashion, subsequent traces of (\ref{diverM}) would imply  
$\m_{\vf}^{\, [k]} = 0$, for $k = n, \, n-1 \, \dots$ and then finally $\m_{\vf} = 0$,
as desired\ft{This is of course true given that the coefficients $\l_k$ 
do not imply
any \emph{identical} cancellations among the 
traces of $\prd M_{\, \vf}$.}.
The form of $\m_{\vf}$ immediately fixes the first
coefficient to be $b_1  =  -  1$,
whereas consistency with (\ref{diverM})
requires
\be
\begin{split}
&\l_{\, k} \, = \, - \, \fr{b_{\, k}}{2\, k\, - \, 1} \, , \\
&b_{\, k\, + \, 1} \, = \, \fr{b_{\, k}}{2\, k\, - \, 1} \, ,
\end{split}
\ee
whose unique solution is 
\be
b_{\, k + 1} \, = \, - \, \fr{1}{(2 \, k \, - \, 1)\,! !} \, \,  . 
\ee
We obtain in this way the complete form of the generalised Fierz-Pauli mass term:
\be \label{mass}
M_{\, \vf} \, = \,  \vf -  \h \, \vf^{\, \pe}  -  \h^{\, 2} \, \vf^{ \, \pe \pe} 
- \fr{1}{3} \, \h^{\, 3} \, \vf^{ \, \pe \pe \pe} - \, \cdots \, - 
\fr{1}{(2\, k \, - \, 3)\, ! !} \, \h^{\, k} \, \vf^{\, [k]} \, - \,  \dots \, .
\ee
Once the equations of motion are reduced to the form
$
\cA_{\, \vf} \, - \, m^2 \, M_{\, \vf} \, = \, 0 \, ,
$
then all traces
of $\vf$ can subsequently shown to vanish, 
thus leading to the conclusion that the Lagrangian equations 
obtained by (\ref{geomlagmass}) imply the system
$(\Box \, - \, m^2) \, \vf \, = \, 0 \, , \ \ 
\prd \vf \, = \, 0 \, , \ \
\vf^{\, \pe} \, = \, 0 \, ,
$
and thus provide a consistent description of massive
higher-spin degrees of freedom. 

\section{Interaction with external sources} \label{sec3}

 One simple possibility to study massive higher-spin fields in a
$d$-dimensional flat background is to deduce their properties from those
of the corresponding massless theory in $d + 1$ dimensions, 
subject to a standard procedure of Kaluza-Klein reduction.
In this fashion, starting  from the massless, unconstrained local Lagrangians
of \cite{fs3, FMS}, it was possible in \cite{FMS, fms08} to compute
the massive propagator for a spin-$s$ field coupled to a 
conserved source. The result is 
\be 
\label{massiveflat} 
(p^2 \, - \, M^2)\, \cJ \cdot \vf \, =
\, \sum_{n=0}^{[\fr{s}{2}]}\, \rho_n\, (d-1,\, s) \, \fr{s!}{n!\, (s - 2n)! \, 2^n}\,
\cJ^{\, [n]}\cdot \cJ^{\, [n]} \, . 
\ee
where the coefficients $\rho_n\, (d-1,\, s)$ are given by
\be
\rho_n\, (d-1,\, s) \, = \, (-1)^n \, \prod_{k =1}^n \, \fr{1}{d - 1 \, + \, 2 \, (s - k -1)} \, .
\ee
 We would like to compare this result with the propagator 
of the theory defined by the Lagrangian
\be 
\cL \, = \,  \12 \, \vf \, \{\cA_{\vf} \, - \, \12 \, \h  \, 
\cA_{\vf}^{\, \pe} \, + \, \h^{\, 2} \, \cB_{\, \vf} \, - \, 
m^{\, 2} \, M_{\, \vf} \}\, - \, \vf \, \cdot \, \cJ \, ,
\ee
where $M_{\, \vf}$ is given by (\ref{mass}), and the field
$\vf$ is coupled to a 
current which we assume to be conserved 
\ft{For a discussion see \cite{fms08}.}.
Under this condition the divergence of the equations
of motion implies the same consequences as for
the free case, and the Lagrangian equation reduces to
\be \label{currenteq}
\cA_{\vf} \, - \, m^{\, 2} \, (\vf -  \h \, \vf^{\, \pe} \, - \, \cdots \, - 
\fr{1}{(2\, k \, - \, 3)\, !!} \, \h^{\, k} \, \vf^{\, [k]} \, \dots) \, = \, \cJ \, .
\ee
Successive traces of this last equation, taking into
account that under the assumed conditions $\cA_{\, \vf}^{\, \pe} = 0$,
give the following system
\be \label{systemexpl}
 \begin{split}
& \vf^{\, \pe} \, + \, \h \, \vf^{\, \pe \pe} \, + \, \h^{\, 2} \, \fr{1}{3}\vf^{\, [3]}\cdots \, + 
\fr{1}{(2\, n \, - \, 3)\, !!} \, \h^{\, n - 1} \, \vf^{\, [n]} \, = \, - \, \fr{\r_{\,1}}{m^{\, 2}} \, 
 \, \cJ^{\, \pe} \, , \\
& \vf^{\, \pe \pe} \, + \, \cdots \, + 
\fr{1}{(2\, n \, - \, 3)\, !!} \, \h^{\, n - 2} \, \vf^{\, [n]} = \, + \, \fr{\r_{\,2}}{m^{\, 2}} \, \cJ^{\, \pe \pe} \, , \\
& \dots  \, \, \dots \, , \\
& \sum_{k = l}^{n} \fr{\h^{\, k - l}}{(2\, l \, - \, 3)\, !!}\vf^{\, [l]} \,  = \, 
(-1)^{\, l} \, \fr{\r_{\, l}}{m^{\, 2}} \, \cJ^{\, [l]} \, , \\
& \dots \, , \\
&\fr{1}{(2\, n \, - \, 3)!!}\,\vf^{\, [n]} \, = \, (-1)^{\, n} \, \fr{\r_{\, n}}{m^{\, 2}} \, \cJ^{\, [n]} \, ,
\end{split}
\ee
where we recall that $n = [\fr{s}{2}]$. It is remarkable that the coefficient of  $\vf^{\, [k]}$ 
is the same in throughout the system, for all $k$. In this sense, 
each of the l.h.s. in (\ref{systemexpl}) really looks like a ``right-truncation''
of the l.h.s. of (\ref{currenteq}). Of course this is not
strictly true, because of the combinatorial factors to be introduced
in order to restore matching between powers of $\h$,
so that, for instance, from the equation for $\cJ^{\, \pe}$ 
we find
\begin{gather} \label{firstrace}
\h \, \sum_{k = 1}^n
\fr{1}{(2\, k \, - \, 3)\, !!} \, \h^{\, k - 1} \, \vf^{\, [k]} \, = \, - \, \h \, \fr{\r_{\,1}}{m^{\, 2}} \, 
 \, \cJ^{\, \pe} \, , \nonumber \\
 \Rightarrow   \\
\sum_{k = 1}^n \, \fr{1}{(2\, k \, - \, 3)\, !!} \, \h^{\, k} \, \vf^{\, [k]} \, = 
\, - \, \h \, \fr{\r_{\,1}}{m^{\, 2}} \, \cJ^{\, \pe} \,+ \, 
\sum_{k = 2}^n \, \fr{1 - k}{(2\, k \, - \, 3)\, !!} \, \h^{\, k} \, \vf^{\, [k]} \, , \nonumber
\end{gather}
which, in its turn, upon substitution in (\ref{currenteq}) gives
\be
\cA \, - \, m^{\, 2} \, \vf \, = \, \cJ \, + \, \r_1 \, \h \, \cJ^{\, \pe} \,
+ \, m^{\, 2} \, \sum_{k = 2}^n \, \fr{k - 1}{(2\, k \, - \, 3)\, !!} \, \h^{\, k} \, \vf^{\, [k]} \, .
\ee
This observation suggests a quicker way to
look for the solution of (\ref{currenteq}): rather than solving
directly for the  $\vf^{\, [k]}$ in terms of
$\cJ^{\, [k + 1]}, \, \cJ^{\, [k + 2]}, \dots $ we shall substitute \emph{the lines}
of (\ref{systemexpl})
in (\ref{currenteq}),  taking care at each step of the corresponding remainder. 
Iterating this procedure one can prove by induction the following relation 
\be
\cA \, - \, m^{\, 2} \, \vf \, = \, \cJ \, + \sum_{k = 1}^l \, \r_{\, k} \, \h^{\, k} \, \cJ^{\, [k]} \, 
+ \, (- 1)^{l + 1} \, m^{\, 2} \, \sum_{k = l + 1}^n \, \fr{{k - 1 \choose l}}{(2\, k \, - \, 3)\, !!} \, \h^{\, k} \, \vf^{\, [k]} \, ,
\ee
where for $l = n$ the remainder is not present, thus 
making it possible to identify
the  projection of the current giving rise to the massive propagator
\be \label{proj}
\cA \, - \, m^{\, 2} \, \vf \, = \, \cJ \, + \, \h \, \r_{\, 1} \, (d - 1, s)\, \cJ^{\, \pe} \, + \,
\h^{\, 2} \, \r_{\, 2}\, (d - 1, s) \, \cJ^{\, \pe \pe} \, + \, \dots \, + \, 
\h^{\, n} \, \r_{\, n}\, (d - 1, s)  \, \cJ^{\, [n]} \, ,
\ee
consistently with (\ref{massiveflat}).

 It might be worth stressing that in the computation of the 
current interaction, under the assumption that the source be conserved, 
the structure of the Einstein tensor (\ref{einst}) plays no role, but
for the main feature that it be divergenceless. This means that the detailed information
about the coefficients of the projector obtained in (\ref{proj}) is entirely encoded in the 
form of the mass term (\ref{mass}), that receives in this way a non-trivial check
on its correctness and uniqueness\ft{See \cite{dario07}, Sec. $4$, for further considerations
on the issue of uniqueness.}. We expect the same mass term (\ref{mass})
to generate also consistent quadratic deformations of the massless geometric
theory generalised to (A)dS backgrounds. The construction of this theory
and the analysis of the corresponding current exchanges,
to be compared with the results recently found in \cite{fms08}, are left for future work.
\begin{acknowledgement}
I would like to thank J. Mourad and A. Sagnotti for discussions and collaboration. 
The present research was supported by the EU contract MRTN-CT-2004-512194.
\end{acknowledgement}

\end{document}